\documentclass[preprintnumbers,amsmath,amssymb]{revtex4}
\usepackage{docs}

\usepackage{graphicx}
\usepackage{dcolumn}

\usepackage{bm}

\begin{document}


\title{Thin shells around traversable wormholes}%
\author{Francisco S. N. Lobo}%
\email{flobo@cosmo.fis.fc.ul.pt} \affiliation{Centro de Astronomia
e Astrof\'{\i}sica da Universidade de Lisboa,\\
Campo Grande, Ed. C8 1749-016 Lisboa, Portugal}












\begin{abstract}

Applying the Darmois-Israel thin shell formalism, we construct
static and dynamic thin shells around traversable wormholes.
Firstly, by applying the cut-and-paste technique we apply a
linearized stability analysis to thin-shell wormholes in the
presence of a generic cosmological constant. We find that for
large positive values of the cosmological constant, i.e., the
Schwarzschild-de Sitter solution, the regions of stability
significantly increase relatively to the Schwarzschild case,
analyzed by Poisson and Visser. Secondly, we construct static thin
shell solutions by matching an interior wormhole solution to a
vacuum exterior solution at a junction surface. In the spirit of
minimizing the usage of exotic matter we analyze the domains in
which the weak and null energy conditions are satisfied at the
junction surface. The characteristics and several physical
properties of the surface stresses are explored, namely, we
determine regions where the sign of tangential surface pressure is
positive and negative (surface tension). An equation governing the
behavior of the radial pressure across the junction surface is
deduced. Specific dimensions of the wormhole, namely, the throat
radius and the junction interface radius, are found by taking into
account the traversability conditions, and estimates for the
traversal time and velocity are also determined.

\end{abstract}

\maketitle



\section{Introduction}

Interest in traversable wormholes, as hypothetical shortcuts in
spacetime, has been rekindled by the classical paper by Morris and
Thorne \cite{Morris}. The subject has also served to stimulate
research in several branches, for instance, the violation of the
energy conditions \cite{VisserBOOK,VisserEC}, closed timelike
curves and the associated difficulties in causality violation
\cite{VisserBOOK,VissEC,LoboCTC}, and superluminal travel
\cite{LoboSLT}, amongst others.

As the violation of the energy conditions is a problematic issue,
depending on one's point of view, it is useful to minimize the
usage of exotic matter. For instance, an elegant way of achieving
this, is to construct a simple class of wormhole solutions using
the cut and paste technique, implemented by Visser
\cite{VisserBOOK,Visser1,Visser2}, in which the exotic matter is
concentrated at the wormhole throat. The surface stress-energy
tensor components of the exotic matter at the throat are
determined, invoking the Darmois-Israel formalism
\cite{Darmois,Israel,Papahamoui}. These thin-shell wormholes are
extremely useful as one may apply a stability analysis for the
dynamical cases, either by choosing specific surface equations of
state \cite{eqstate1,eqstate2,eqstate3}, or by considering a
linearized stability analysis around a static solution
\cite{Poisson,Lobolinear}, in which a parametrization of the
stability of equilibrium is defined, so that one does not have to
specify a surface equation of state. In \cite{Lobolinear}, the
generalization of the Poisson-Visser linearized analysis
\cite{Poisson} for thin-shell wormholes was done in the presence
of a non-vanishing cosmological constant. It was found that for
large positive values of $\Lambda$, i.e., the Schwarzschild-de
Sitter solution, the regions of stability significantly increase
relatively to the Schwarzschild case analyzed by Poisson and
Visser.

As an alternative to these thin-shell wormholes, one may also
consider that the exotic matter is distributed from the throat to
a radius $a$, where the solution is matched to an exterior vacuum
spacetime. Several simple cases were analyzed in \cite{Morris},
but one may invoke the Darmois-Israel formalism to consider a
broader class of solutions. Thus, the thin shell confines the
exotic matter threading the wormhole to a finite region, with a
delta-function distribution of the stress-energy tensor on the
junction surface. The motivation of this analysis resides in
minimizing the usage of exotic matter threading the interior
wormhole solution, which in principle may be made arbitrarily
small \cite{VKD,Viss}, and imposing that the surface stresses of
the thin shell obey the energy conditions. One may generalize and
systematize \cite{Lobo} the particular case of a matching with a
constant redshift function and a null surface energy density on
the junction boundary, studied in \cite{LLQ}. A similar analysis
for the plane symmetric case, with a negative cosmological
constant, is done in \cite{LL}. The plane symmetric traversable
wormhole is a natural extension of the topological black hole
solutions found by Lemos \cite{lemos1,lemos2,lemos3,zanchin}, upon
addition of exotic matter. These plane symmetric wormholes may be
viewed as domain walls connecting different universes. They may
have planar topology, and upon compactification of one or two
coordinates, cylindrical topology or toroidal topology,
respectively.

This paper is organized as follows. In Section II the
Darmois-Israel thin shell formalism is briefly reviewed. In
Section III, a linearized stability analysis is studied in the
presence of a generic cosmological constant \cite{Lobolinear}. In
Section IV we match an interior wormhole spacetime to an exterior
vacuum solution and find specific regions where the energy
conditions at the junction are obeyed \cite{Lobo}; we also analyze
the physical properties and characteristics of the surface
stresses, namely, we find domains where the tangential surface
pressure is positive or negative (surface tension); we deduce an
expression governing the behavior of the radial pressure across
the junction boundary; and finally, we determine specific
dimensions of the wormhole by taking into account the
traversability conditions and find estimates for the traversal
time and velocity. Finally, in Section V, we conclude.

\section{Overview of the Darmois-Israel formalism}

Consider two distinct spacetime manifolds, ${\cal M_+}$ and ${\cal
M_-}$, with metrics given by $g_{\mu \nu}^+(x^{\mu}_+)$ and
$g_{\mu \nu}^-(x^{\mu}_-)$, in terms of independently defined
coordinate systems $x^{\mu}_+$ and $x^{\mu}_-$. The manifolds are
bounded by hypersurfaces $\Sigma_+$ and $\Sigma_-$, respectively,
with induced metrics $g_{ij}^+$ and $g_{ij}^-$. The hypersurfaces
are isometric, i.e., $g_{ij}^+(\xi)=g_{ij}^-(\xi)=g_{ij}(\xi)$, in
terms of the intrinsic coordinates, invariant under the isometry.
A single manifold ${\cal M}$ is obtained by gluing together ${\cal
M_+}$ and ${\cal M_-}$ at their boundaries, i.e., ${\cal M}={\cal
M_+}\cup {\cal M_-}$, with the natural identification of the
boundaries $\Sigma=\Sigma_+=\Sigma_-$.

The three holonomic basis vectors $e_{(i)}=\partial /\partial
\xi^i$ tangent to $\Sigma$ have the following components
$e^{\mu}_{(i)}|_{\pm}=\partial x_{\pm}^{\mu}/\partial \xi^i$,
which provide the induced metric on the junction surface by the
following scalar product
\begin{equation}
g_{ij}=e_{(i)}\cdot e_{(j)}=g_{\mu
\nu}e^{\mu}_{(i)}e^{\nu}_{(j)}|_{\pm}.
\end{equation}

We shall consider a timelike junction surface $\Sigma$, defined by
the parametric equation of the form $f(x^{\mu}(\xi^i))=0$. The
unit normal $4-$vector, $n^{\mu}$, to $\Sigma$ is defined as
\begin{equation}\label{defnormal}
n_{\mu}=\pm \,\left |g^{\alpha \beta}\,\frac{\partial f}{\partial
x ^{\alpha}} \, \frac{\partial f}{\partial x ^{\beta}}\right
|^{-1/2}\;\frac{\partial f}{\partial x^{\mu}}\,,
\end{equation}
with $n_{\mu}\,n^{\mu}=+1$ and $n_{\mu}e^{\mu}_{(i)}=0$. The
Israel formalism requires that the normals point from ${\cal M_-}$
to ${\cal M_+}$ \cite{Israel}.

The extrinsic curvature, or the second fundamental form, is
defined as $K_{ij}=n_{\mu;\nu}e^{\mu}_{(i)}e^{\nu}_{(j)}$, or
\begin{eqnarray}\label{extrinsiccurv}
K_{ij}^{\pm}=-n_{\mu} \left(\frac{\partial ^2 x^{\mu}}{\partial
\xi ^{i}\,\partial \xi ^{j}}+\Gamma ^{\mu \pm}_{\;\;\alpha
\beta}\;\frac{\partial x^{\alpha}}{\partial \xi ^{i}} \,
\frac{\partial x^{\beta}}{\partial \xi ^{j}} \right) \,.
\end{eqnarray}
Note that for the case of a thin shell $K_{ij}$ is not continuous
across $\Sigma$, so that for notational convenience, the
discontinuity in the second fundamental form is defined as
$\kappa_{ij}=K_{ij}^{+}-K_{ij}^{-}$.

Now, the Lanczos equations follow from the Einstein equations for
the hypersurface, and are given by
\begin{equation}
S^{i}_{\;j}=-\frac{1}{8\pi}\,(\kappa ^{i}_{\;j}-\delta
^{i}_{\;j}\kappa ^{k}_{\;k})  \,,
\end{equation}
where $S^{i}_{\;j}$ is the surface stress-energy tensor on
$\Sigma$.

The first contracted Gauss-Kodazzi equation or the ``Hamiltonian"
constraint
\begin{eqnarray}
G_{\mu \nu}n^{\mu}n^{\nu}=\frac{1}{2}\,(K^2-K_{ij}K^{ij}-\,^3R)\,,
\end{eqnarray}
with the Einstein equations provide the evolution identity
\begin{eqnarray}
S^{ij}\overline{K}_{ij}=-\left[T_{\mu
\nu}n^{\mu}n^{\nu}-\Lambda/8\pi \right]\,.
\end{eqnarray}
The convention $\left[X \right]\equiv X^+|_{\Sigma}-X^-|_{\Sigma}$
and $\overline{X} \equiv (X^+|_{\Sigma}+X^-|_{\Sigma})/2$ is used.

The second contracted Gauss-Kodazzi equation or the ``ADM"
constraint
\begin{eqnarray}
G_{\mu \nu}e^{\mu}_{(i)}n^{\nu}=K^j_{i|j}-K,_{i}\,,
\end{eqnarray}
with the Lanczos equations gives the conservation identity
\begin{eqnarray}
S^{i}_{j|i}=\left[T_{\mu \nu}e^{\mu}_{(j)}n^{\nu}\right]\,.
\end{eqnarray}

In particular, considering spherical symmetry considerable
simplifications occur, namely $\kappa ^{i}_{\;j}={\rm diag}
\left(\kappa ^{\tau}_{\;\tau},\kappa ^{\theta}_{\;\theta},\kappa
^{\theta}_{\;\theta}\right)$. The surface stress-energy tensor may
be written in terms of the surface energy density, $\sigma$, and
the surface pressure, $p$, as $S^{i}_{\;j}={\rm
diag}(-\sigma,p,p)$. The Lanczos equations then reduce to
\begin{eqnarray}
\sigma &=&-\frac{1}{4\pi}\,\kappa ^{\theta}_{\;\theta} \,,\label{sigma} \\
p &=&\frac{1}{8\pi}(\kappa ^{\tau}_{\;\tau}+\kappa
^{\theta}_{\;\theta}) \,. \label{surfacepressure}
\end{eqnarray}

\section{Cut and paste technique:
Thin-shell wormholes with a cosmological constant}

In this section we shall construct a class of wormhole solutions,
in the presence of a cosmological constant, using the
cut-and-paste technique. Consider the unique spherically symmetric
vacuum solution, i.e.,
\begin{eqnarray}
ds^2=-\left(1-\frac{2M}{r}-\frac{\Lambda}{3}r^2
\right)\,dt^2+\left(1-\frac{2M}{r}-\frac{\Lambda}{3}r^2
\right)^{-1}\,dr^2 +r^2\,(d\theta ^2+\sin ^2{\theta}\, d\phi ^2)
\label{metricvacuumlambda}.
\end{eqnarray}
If $\Lambda >0$, the solution is denoted by the Schwarzschild-de
Sitter metric. For $\Lambda <0$, we have the Schwarzschild-anti de
Sitter metric, and of course the specific case of $\Lambda =0$ is
reduced to the Schwarzschild solution. Note that the metric $(1)$
is not asymptotically flat as $r \rightarrow \infty$. Rather, it
is asymptotically de Sitter, if $\Lambda >0$, or asymptotically
anti-de Sitter, if $\Lambda <0$. But, considering low values of
$\Lambda$, the metric is almost flat in the range $M \ll r \ll
1/\sqrt{\Lambda}$. For values below this range, the effects of $M$
dominate, and for values above the range, the effects of $\Lambda$
dominate, as for large values of the radial coordinate the
large-scale structure of the spacetime must be taken into account.

The specific case of $\Lambda =0$ is reduced to the Schwarzschild
solution, with a black hole event horizon at $r_b=2M$. Consider
the Schwarzschild-de Sitter spacetime, $\Lambda
>0$. If $0<9\Lambda M^2<1$, the factor $g(r)=(1-2M/r-\Lambda r^2/3)$
possesses two positive real roots, $r_b$ and $r_c$, corresponding
to the black hole and the cosmological event horizons of the de
Sitter spacetime, respectively, given by
\begin{eqnarray}
r_b&=&2 \Lambda ^{-1/2} \, \cos(\alpha/3)    \label{root1}  \,, \\
r_c&=&2 \Lambda ^{-1/2} \, \cos(\alpha/3+4\pi/3)    \label{root2}
\,,
\end{eqnarray}
where $\cos \alpha \equiv -3M \Lambda^{1/2}$, with $\pi < \alpha
<3\pi/2$. In this domain we have $2M<r_b<3M$ and $r_c>3M$.

For the Schwarzschild-anti de Sitter metric, with $\Lambda <0$,
the factor $g(r)=\left(1-2M/r+|\Lambda |r^2/3 \right)$ has only
one real positive root, $r_b$, given by
\begin{eqnarray}
r_b=\left(\frac{3M}{|\Lambda|}\right)^{1/3}\left(\sqrt[3]{1+\sqrt{1+\frac{1}{9|\Lambda|M^2}}}
+\sqrt[3]{1-\sqrt{1+ \frac{1}{9|\Lambda|M^2}}}\;\right)
\label{adsbhole} ,
\end{eqnarray}
corresponding to a black hole event horizon, with $0<r_b<2M$.

\subsection{The cut-and-paste construction}

Given this, we may construct a wormhole solution, using the
cut-and-paste technique \cite{VisserBOOK,Visser1,Visser2}.
Consider two vacuum solutions with $\Lambda$ and remove from each
spacetime the region described by
\begin{equation}
\Omega_{\pm}\equiv \left \{r_{\pm}\leq a| \,a >r_b \right \} \,,
\end{equation}
where $a$ is a constant and $r_b$ is the black hole event horizon,
corresponding to the Schwarzschild-de Sitter and
Schwarzschild-anti de Sitter solutions, equation (\ref{root1}) and
equation (\ref{adsbhole}), respectively. The removal of the
regions results in two geodesically incomplete manifolds, with
boundaries given by the following timelike hypersurfaces
\begin{equation}
\Sigma_{\pm}\equiv \left \{r_{\pm}= a| \,a > r_b \right \} \,.
\end{equation}
Identifying these two timelike hypersurfaces,
$\Sigma_{+}=\Sigma_{-}$, results in a geodesically complete
manifold, with two regions connected by a wormhole and the
respective throat situated at $\Sigma$. The wormhole connects two
regions, asymptotically de Sitter or anti-de Sitter, for $\Lambda
>0$ and $\Lambda <0$, respectively.

The intrinsic metric at $\Sigma$ is given by
\begin{equation}
ds^2_{\Sigma}=-d\tau ^2 +a^2(\tau)\,(d\theta ^2+\sin ^2{\theta}\,
d\phi ^2)\,,
\end{equation}
where $\tau$ is the proper time as measured by a comoving observer
on the wormhole throat.

\subsection{The surface stresses}

The imposition of spherical symmetry is sufficient to conclude
that there is no gravitational radiation, independently of the
behavior of the wormhole throat. The position of the throat is
given by
$x^{\mu}(\tau,\theta,\phi)=(t(\tau),a(\tau),\theta,\phi)$, and the
respective $4$-velocity is
\begin{equation}
U^{\mu}=\left(\frac{\sqrt{1-2M/a-\Lambda a^2/3+\dot{a}^2}}
{1-2M/a-\Lambda a^2/3},\dot{a},0,0 \right)  \,,
\end{equation}
where the overdot denotes a derivative with respect to $\tau$.

The unit normal to the throat may be determined by equation
(\ref{defnormal}) or by the contractions, $U^{\mu}n_{\mu}=0$ and
$n^{\mu}n_{\mu}=+1$, and is given by
\begin{equation}
n^{\mu}=\left(\frac{\dot{a}} {1-2M/a-\Lambda
a^2/3},\sqrt{1-2M/a-\Lambda a^2/3+\dot{a}^2},0,0 \right)
\label{normal} \,.
\end{equation}

Using equation (\ref{extrinsiccurv}), equation
(\ref{metricvacuumlambda}) and equation (\ref{normal}), the
non-trivial components of the extrinsic curvature are given by
\begin{eqnarray}
K^{\theta\;\pm}_{\;\,\theta}&=& \pm
\frac{1}{a}\sqrt{1-2M/a-\Lambda a^2/3+\dot{a}^2} \;,
    \label{Kplustheta}\\
K ^{\tau \;\pm}_{\;\,\tau}&=& \pm \frac{M/a^2-\Lambda a/3+
\ddot{a}} {\sqrt{1-2M/a-\Lambda a^2/3+\dot{a}^2}} \;,
    \label{Kplustautau}
\end{eqnarray}
Thus, the Einstein field equations, equations
(\ref{sigma})-(\ref{surfacepressure}), with equations
(\ref{Kplustheta})-(\ref{Kplustautau}), then provide us with the
following surface stresses
\begin{eqnarray}
\sigma&=&-\frac{1}{2\pi a} \sqrt{1-2M/a-\Lambda a^2/3+\dot{a}^2}
  \label{surfenergy1}   \,,\\
p&=&\frac{1}{4\pi a} \;\frac{1-M/a-2\Lambda
a^2/3+\dot{a}^2+a\ddot{a}}{\sqrt{1-2M/a-\Lambda a^2/3+\dot{a}^2}}
\label{surfpressure1}  \,.
\end{eqnarray}

We also verify that the above equations imply the conservation of
the surface stress-energy tensor
\begin{equation}
\dot{\sigma}=-2 \left(\sigma +p \right) \frac{\dot{a}}{a}
   \label{conservationenergy}
\end{equation}
or
\begin{equation}
\frac{d\left(\sigma A \right)}{d\tau}+p\,\frac{dA}{d\tau}=0 \,,
\end{equation}
where $A=4\pi a^2$ is the area of the wormhole throat. The first
term represents the variation of the internal energy of the
throat, and the second term is the work done by the throat's
internal forces.

\subsection{Linearized stability analysis}

Equation (\ref{surfenergy1}) may be recast into the following
dynamical form
\begin{equation}
\dot{a}^2-\frac{2M}{a}-\frac{\Lambda}{3} a^2- \left(2\pi \sigma a
\right)^2=-1  \,,  \label{motionofthroat}
\end{equation}
which determines the motion of the wormhole throat. Considering an
equation of state of the form, $p=p(\sigma)$, the energy
conservation, equation (\ref{conservationenergy}), can be
integrated to yield
\begin{equation}
\ln(a)=-\frac{1}{2}\int \frac{d\sigma}{\sigma +p(\sigma)}  \,.
\end{equation}
This result formally inverted to provide $\sigma=\sigma(a)$, can
be finally substituted into equation (\ref{motionofthroat}). The
latter can also be written as $\dot{a}^2=-V(a)$, with the
potential defined as
\begin{equation}
V(a)=1-\frac{2M}{a}-\frac{\Lambda}{3} a^2- \left(2\pi \sigma a
\right)^2     \label{defpotential}  \,.
\end{equation}

One may explore specific equations of state, but following the
Poisson-Visser analysis \cite{Poisson}, we shall consider a linear
perturbation around a static solution with radius $a_0$. The
respective values of the surface energy density and the surface
pressure, at $a_0$, are given by
\begin{eqnarray}
\sigma_0&=&-\frac{1}{2\pi a_0} \sqrt{1-2M/a_0-\Lambda
a_0^2/3}   \label{stablesigma} \,,\\
p_0&=&\frac{1}{4\pi a_0}\;\frac{1-M/a_0-2\Lambda
a_0^2/3}{\sqrt{1-2M/a_0-\Lambda a_0^2/3}} \,.
\label{stablepressure}
\end{eqnarray}
One verifies that the surface energy density is always negative,
implying the violation of the weak and dominant energy conditions.
One may verify that the null and strong energy conditions are
satisfied for $a_0 \leq 3M$ for a generic $\Lambda$, for a generic
cosmological constant \cite{Lobolinear}.

Linearizing around the stable solution at $a=a_0$, we consider a
Taylor expansion of $V(a)$ around $a_0$ to second order, which
provides
\begin{equation}
 V(a)=V(a_0)+V'(a_0)(a-a_0)+\frac{1}{2}\,V''(a_0)(a-a_0)^2+O
\left[(a-a_0)^3 \right] \label{Taylorexpansion} ,
\end{equation}
where the prime denotes a derivative with respect to $a$, $d/da$.

Define the parameter $\eta(\sigma)=dp/d\sigma=p'/\sigma'$. The
physical interpretation of $\eta$ is discussed in \cite{Poisson},
and $\sqrt{\eta}$ is normally interpreted as the speed of sound.
Evaluated at the static solution, at $a=a_0$, using equations
(\ref{stablesigma})-(\ref{stablepressure}), we readily find
$V(a_0)=0$ and $V'(a_0)=0$, with $V''(a_0)$ given by
\begin{equation}
V''(a_0)=-\frac{2}{a_0^2} \Bigg[
\frac{2M}{a_0}+\frac{\Lambda}{3}a_0^2+\frac{\left(M/a_0-\Lambda
a_0^2/3 \right)^2}{1-2M/a_0-\Lambda a_0^2/3}+(1+2\eta_0)
\left(1-\frac{3M}{a_0} \right) \Bigg],
\end{equation}
where $\eta_0=\eta (\sigma_0)$.

The potential $V(a)$, equation (\ref{Taylorexpansion}), is reduced
to
\begin{equation}
V(a)=\frac{1}{2}\,V''(a_0)(a-a_0)^2+O \left[(a-a_0)^3 \right]  \,,
\end{equation}
so that the equation of motion for the wormhole throat presents
the following form
\begin{equation}
\dot{a}^2=-\frac{1}{2}V''(a_0)(a-a_0)^2+O \left[(a-a_0)^3 \right]
\,,
\end{equation}
to the order of approximation considered. If $V''(a_0)<0$ is
verified, then the potential $V(a_0)$ has a local maximum at
$a_0$, where a small perturbation in the wormhole throat's radius
will provoke an irreversible contraction or expansion of the
throat. Thus, the solution is stable if and only if $V(a_0)$ has a
local minimum at $a_0$ and $V''(a_0)>0$, i.e.,
\begin{equation}
\eta_0 \left(1-\frac{3M}{a_0} \right) <
-\frac{1-3M/a_0+3M^2/a_0^2-\Lambda Ma_0}{2 \left(1-2M/a_0-\Lambda
a_0^2/3 \right)}
   \label{inequality}   \,.
\end{equation}

We need to analyze equation (\ref{inequality}) for several cases.
The right hand side of equation (\ref{inequality}) is always
negative \cite{Lobolinear}, while the left hand side changes sign
at $a_0=3M$. Thus, one deduces that the stability regions are
dictated by the following inequalities
\begin{eqnarray}
\eta_0<-\frac{1-3M/a_0+3M^2/a_0^2-\Lambda
Ma_0}{2\left(1-2M/a_0-\Lambda a_0^2/3 \right)\left(1-3M/a_0
\right)}\,,\qquad& \,a_0>3M         \label{eta01}
    \\
\eta_0>-\frac{1-3M/a_0+3M^2/a_0^2-\Lambda
Ma_0}{2\left(1-2M/a_0-\Lambda a_0^2/3 \right)\left(1-3M/a_0
\right)}\,,\qquad& \,a_0<3M .      \label{eta02}
\end{eqnarray}

One may analyze several cases.

\subsubsection{1. De Sitter spacetime.}

For the de Sitter spacetime, with $M=0$ and $\Lambda >0$, equation
(\ref{inequality}) reduces to
\begin{equation}
\eta_0<-\frac{1}{2\left(1-\Lambda a_0^2/3 \right)} \,, \;\;\;\;
{\rm for}\;\;0<a_0<\sqrt{3/\Lambda} \,.
\end{equation}
The stability region is depicted in the left plot of figure $1$.

\subsubsection{2. Anti-de Sitter spacetime.}

For the anti-de Sitter spacetime, with $M=0$ and $\Lambda <0$,
equation (\ref{inequality}) gives
\begin{equation}
\eta_0<-\frac{1}{2\left(1+|\Lambda| \,a_0^2/3 \right)} \,,
\;\;\;\; {\rm for}\;\;a_0>0   \label{adsconstraint}\,,
\end{equation}
and the respective stability region is depicted in the right plot
of figure $1$.

\begin{figure}[h]

  \centering
  \includegraphics[width=2.4in]{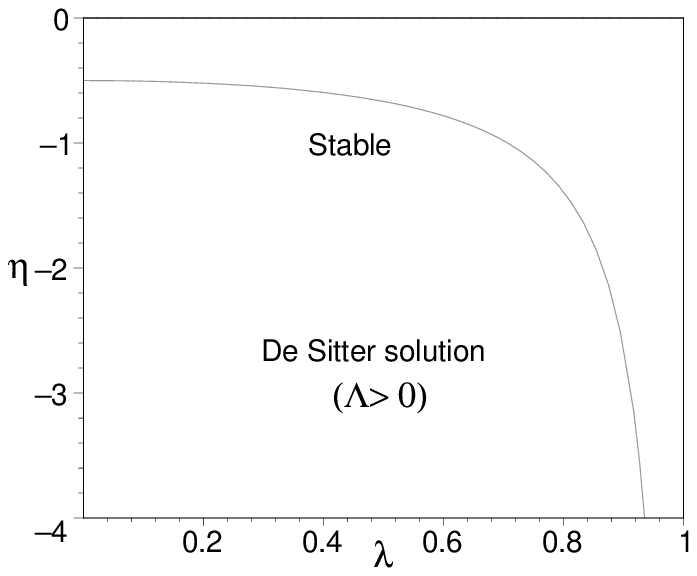}
  \hspace{0.4in}
  \includegraphics[width=2.4in]{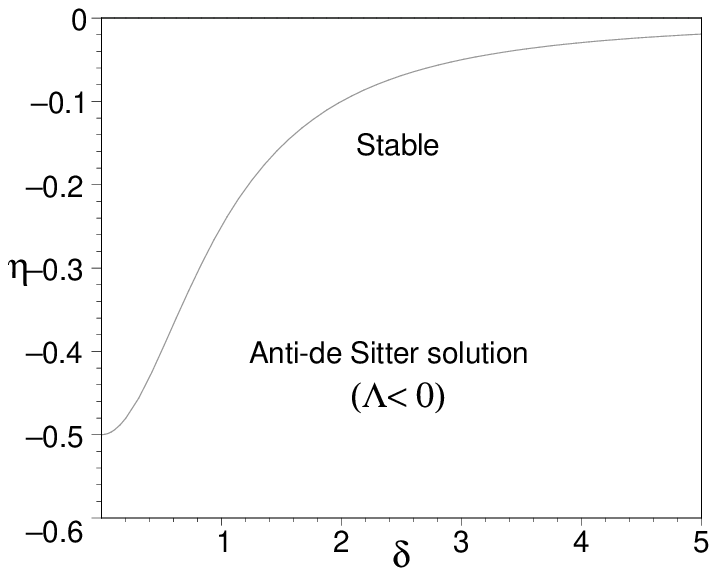}
  \caption{We have defined $\lambda = a_0/(3/\Lambda)^{1/2}$ and
  $\delta = a_0/(3/|\Lambda|)^{1/2}$, respectively.
  The regions of stability are depicted in the graphs,
  below the curves for the de-Sitter and the anti-de Sitter solutions, respectively.}
\end{figure}

\subsubsection{3. Schwarzschild spacetime.}

This is the particular case of the Poisson-Visser analysis
\cite{Poisson}, with $\Lambda=0$, which reduces to
\begin{eqnarray}
\eta_0<-\frac{1-3M/a_0+3M^2/a_0^2}{2\left(1-2M/a
\right)\left(1-3M/a_0 \right)}\,, \qquad& a_0>3M \,,  \\
\eta_0>-\frac{1-3M/a_0+3M^2/a_0^2}{2\left(1-2M/a
\right)\left(1-3M/a_0 \right)}\,, \qquad& a_0<3M \,.
\end{eqnarray}
The stability regions are shown in the left plot of figure $2$.

\subsubsection{4. Schwarzschild-anti de Sitter spacetime.}

For the Schwarzschild-anti de Sitter spacetime, with $\Lambda <0$,
we have
\begin{eqnarray}
\eta_0<-\frac{1-3M/a_0+3M^2/a_0^2+|\Lambda| \,
Ma_0}{2\left(1-2M/a_0+|\Lambda|\, a_0^2/3 \right)\left(1-3M/a_0
\right)}  \,, \qquad& a_0>3M     \label{ads1}
    \\
\eta_0>-\frac{1-3M/a_0+3M^2/a_0^2+|\Lambda| \,
Ma_0}{2\left(1-2M/a_0+|\Lambda|\, a_0^2/3 \right)\left(1-3M/a_0
\right)}\,, \qquad& a_0<3M        \label{ads2}
\end{eqnarray}

The regions of stability are depicted in the right plot of figure
$2$, considering the value $9|\Lambda |M^2=0.9$. In this case, the
black hole event horizon is given by $r_b \simeq 1.8\,M$. We
verify that the regions of stability decrease, relatively to the
Schwarzschild case.

\begin{figure}[h]
  \centering
  \includegraphics[width=2.4in]{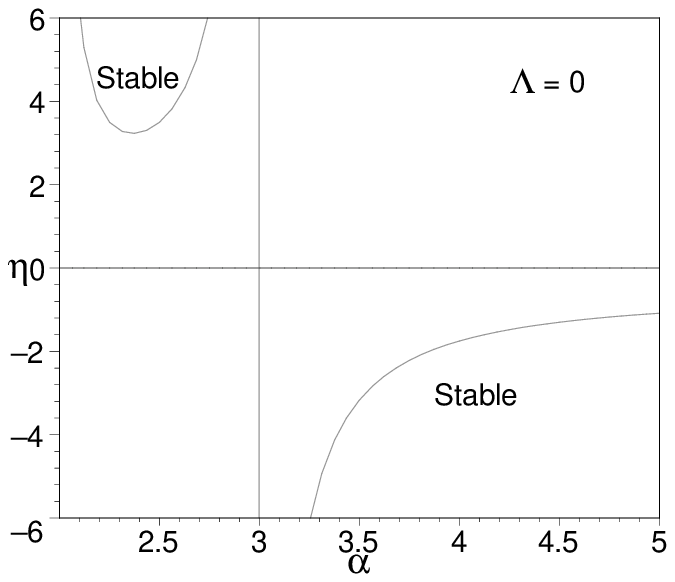}
  \hspace{0.4in}
  \includegraphics[width=2.4in]{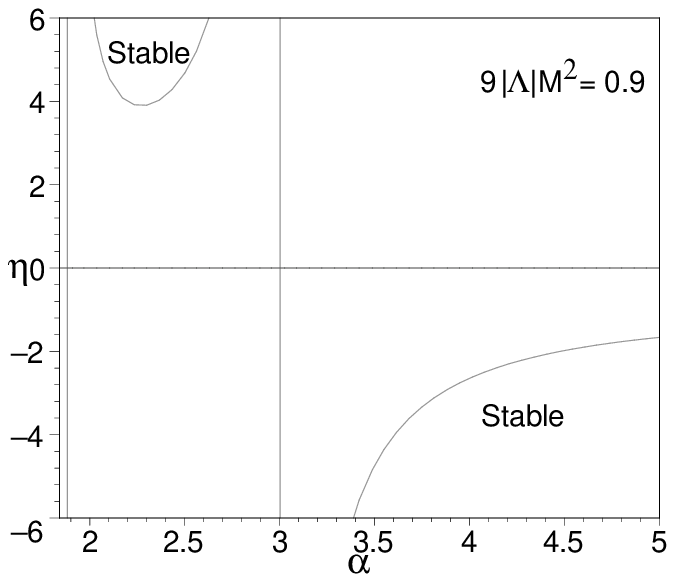}
  \caption{We have defined $\alpha =a_0/M$.
  The regions of stability are depicted for the
  Schwarzschild and the Schwarzschild-anti de Sitter solutions, respectively.
  Imposing the value $9|\Lambda |M^2=0.9$ in the Schwarzschild-anti de Sitter case,
  we verify that the stability regions decrease relatively to the
  Schwarzschild solution.}
\end{figure}

\subsubsection{5. Schwarzschild-de Sitter spacetime.}

For the Schwarzschild-de Sitter spacetime, with $\Lambda
>0$, we have
\begin{eqnarray}
\eta_0<-\frac{1-3M/a_0+3M^2/a_0^2-\Lambda
Ma_0}{2\left(1-2M/a_0-\Lambda a_0^2/3 \right)\left(1-3M/a_0
\right)}   \,,    \qquad&   \,a_0>3M
    \\
\eta_0>-\frac{1-3M/a_0+3M^2/a_0^2-\Lambda
Ma_0}{2\left(1-2M/a_0-\Lambda a_0^2/3 \right)\left(1-3M/a_0
\right)}  \,,    \qquad&     \,a_0<3M .
\end{eqnarray}

The regions of stability are depicted in figure $3$ for increasing
values of $9\Lambda M^2$. In particular, for $9\Lambda M^2=0.7$
the black hole and cosmological horizons are given by $r_b \simeq
2.33\,M$ and $r_c =4.71\,M$, respectively. Thus, only the interval
$2.33 < a_0/M < 4.71$ is taken into account, as shown in the range
of the respective plot. Analogously, for $9\Lambda M^2=0.9$, we
find $r_b \simeq 2.56\,M$ and $r_c =3.73\,M$. Therefore only the
range within the interval $2.56 < a_0/M < 3.73$ corresponds to the
stability regions, also shown in the respective plot.

We verify that for large values of $\Lambda$, or large $M$, the
regions of stability are significantly increased, relatively to
the $\Lambda =0$ case.

\begin{figure}[h]
\centering
  \includegraphics[width=2.4in]{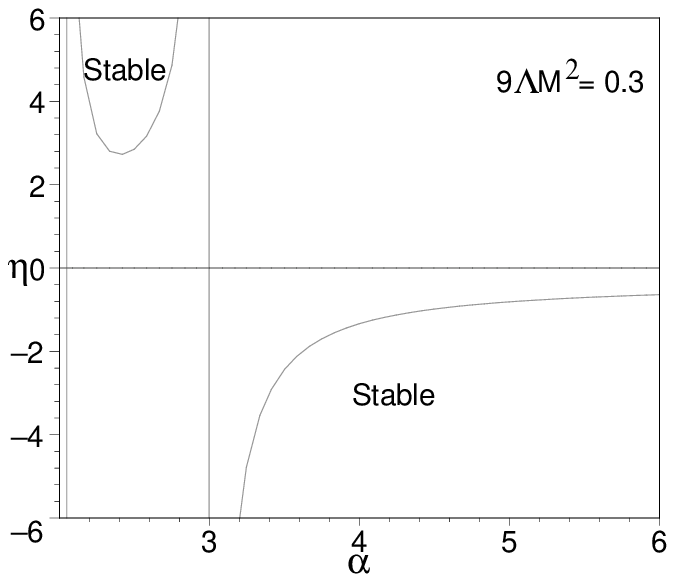}
  \hspace{0.4in}
  \includegraphics[width=2.4in]{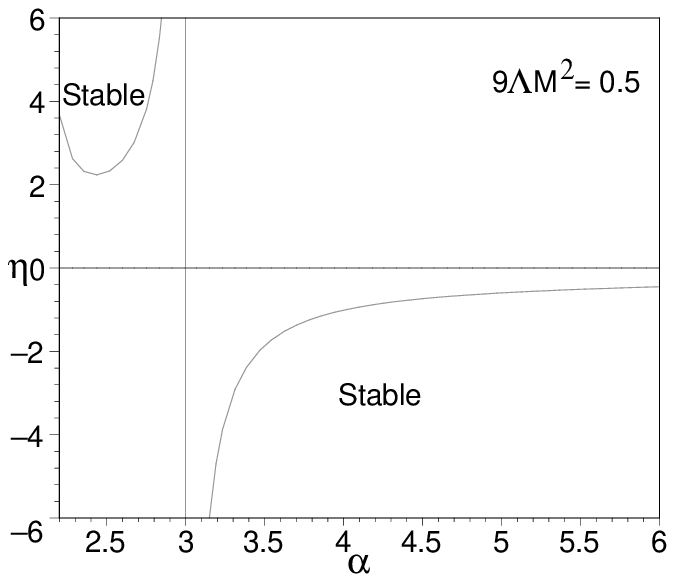}
  \centering
  \includegraphics[width=2.4in]{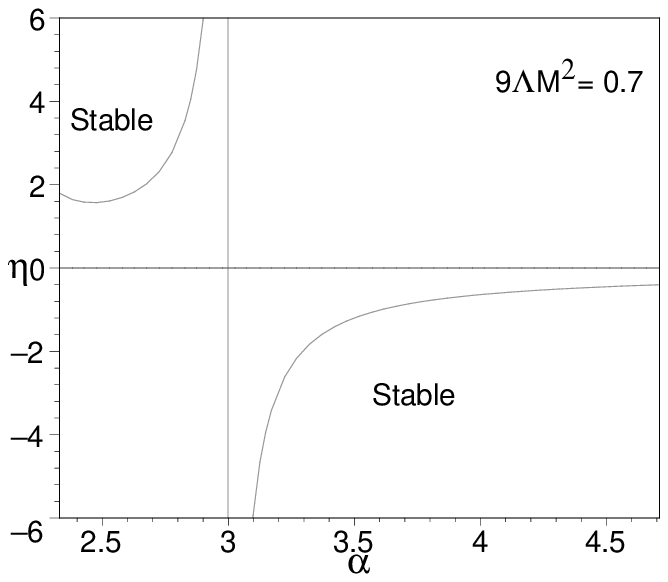}
  \hspace{0.4in}
  \includegraphics[width=2.4in]{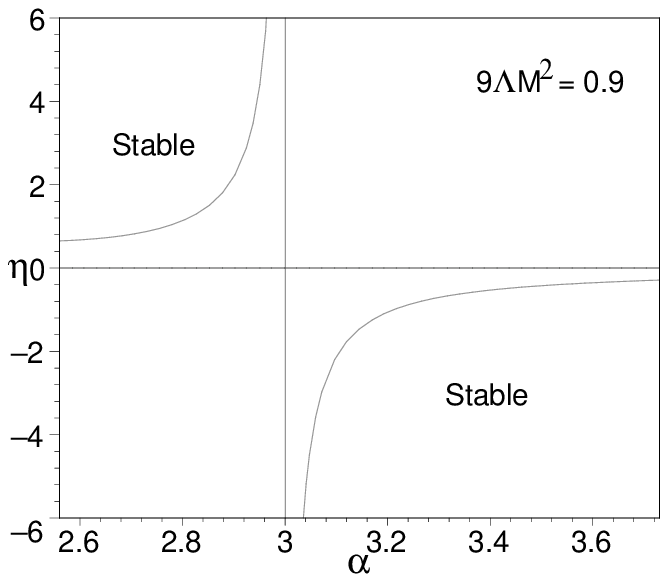}
  \caption{We have defined $\alpha =a_0/M$.
  The regions of stability for the Schwarzschild-de Sitter solution,
  imposing $9\Lambda M^2=0.3$, $9\Lambda M^2=0.5$, $9\Lambda M^2=0.7$
  and $9\Lambda M^2=0.9$, respectively.
  The regions of stability are significantly increased, relatively
  to the $\Lambda =0$ case, for increasing values of $9\Lambda M^2$.}
\end{figure}

\section{A thin shell around a traversable wormhole}

As an alternative to the thin-shell wormhole, constructed using
the cut-and-paste technique analyzed above, one may also consider
that the exotic matter is distributed from the throat $r_0$ to a
radius $a$, where the solution is matched to an exterior vacuum
spacetime. Thus, the thin shell confines the exotic matter
threading the wormhole to a finite region, with a delta-function
distribution of the stress-energy tensor on the junction surface.

We shall match the interior wormhole solution \cite{Morris}
\begin{equation}
ds^2=-e ^{2\Phi(r)}\,dt^2+\frac{dr^2}{1- b(r)/r}+r^2 \,(d\theta
^2+\sin ^2{\theta} \, d\phi ^2) \label{metricwormhole}  \,,
\end{equation}
to the exterior vacuum solution
\begin{eqnarray}
ds^2=-\left(1-\frac{2M}{r}-\frac{\Lambda}{3}r^2
\right)\,dt^2+\left(1-\frac{2M}{r}-\frac{\Lambda}{3}r^2
\right)^{-1}\,dr^2 +r^2\,(d\theta ^2+\sin ^2{\theta}\, d\phi ^2)
\,,
\end{eqnarray}
at a junction surface, $\Sigma$. Thus, the intrinsic metric to
$\Sigma$ is given by
\begin{equation}
ds^2_{\Sigma}=-d\tau^2 + a^2 \,(d\theta ^2+\sin
^2{\theta}\,d\phi^2)  \,.
\end{equation}
Note that the junction surface, $r=a$, is situated outside the
event horizon, i.e., $a>r_b$, to avoid a black hole solution.

Thus, using equation (\ref{extrinsiccurv}), the non-trivial
components of the extrinsic curvature are given by
\begin{eqnarray}
K ^{\tau \;+}_{\;\;\tau}&=&\frac{\frac{M}{a^2}-
\frac{\Lambda}{3}a}{\sqrt{1-\frac{2M}{a}-\frac{\Lambda}{3}a^2}}
\;,  \label{Kplustautau2}\\
K ^{\tau \;-}_{\;\;\tau}&=&\Phi'(a)\sqrt{1-\frac{b(a)}{a}}  \;,
\label{Kminustautau2}
\end{eqnarray}
and
\begin{eqnarray}
K ^{\theta \;+}_{\;\;\theta}&=&\frac{1}{a}\sqrt{1-\frac{2M}{a}-
\frac{\Lambda}{3}a^2}\;,  \label{Kplustheta2}\\
K ^{\theta \;-}_{\;\;\theta}&=&\frac{1}{a}\sqrt{1-\frac{b(a)}{a}}
\;.  \label{Kminustheta2}
\end{eqnarray}

The Einstein equations, equations
(\ref{sigma})-(\ref{surfacepressure}), with the extrinsic
curvatures, equations (\ref{Kplustautau2})-(\ref{Kminustheta2}),
then provide us with the following expressions for the surface
stresses
\begin{eqnarray}
\sigma&=&-\frac{1}{4\pi a} \left(\sqrt{1-\frac{2M}{a}-
\frac{\Lambda}{3}a^2}- \sqrt{1-\frac{b(a)}{a}} \, \right)
    \label{surfenergy2}   ,\\
p&=&\frac{1}{8\pi a} \left(\frac{1-\frac{M}{a}
-\frac{2\Lambda}{3}a^2}{\sqrt{1-\frac{2M}{a}-\frac{\Lambda}{3}a^2}}-
\zeta \, \sqrt{1-\frac{b(a)}{a}} \, \right)
    \label{surfpressure2}    ,
\end{eqnarray}
with $\zeta=1+a\Phi'(a)$. If the surface stress-energy terms are
null, the junction is denoted as a boundary surface. If surface
stress terms are present, the junction is called a thin shell,
which is represented in figure $4$.

\begin{figure}[h]
  \centering
  \includegraphics[width=3.2in]{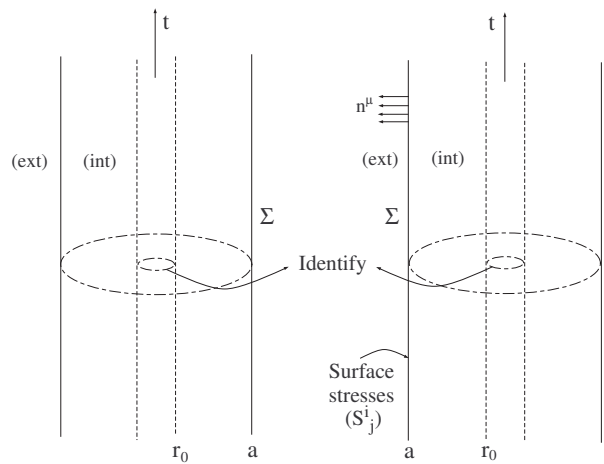}
  \caption{Two copies of static timelike hypersurfaces, $\Sigma$, embedded in
  asymptotic regions, separating an interior wormhole solution from
  an exterior vacuum spacetime. Both copies are identified at the wormhole
  throat, $r_0$. The surface stresses reside on $\Sigma$, and members of the
  normal vector field, $n^{\mu}$, are shown.}
\end{figure}

The surface mass of the thin shell is given by
\begin{eqnarray}\label{shellmass}
M_{\rm shell}=4\pi a^2 \sigma
=a\left(\sqrt{1-\frac{b(a)}{a}}-\sqrt{1-\frac{2M}{a}-
\frac{\Lambda}{3}a^2} \, \right)  .
\end{eqnarray}
One may interpret $M$ as the total mass of the system, in this
case being the total mass of the wormhole in one asymptotic
region. Thus, solving equation (\ref{shellmass}) for $M$, we
finally have
\begin{equation}\label{totalmass}
M=\frac{b(a)}{2}+M_{\rm
shell}\left(\sqrt{1-\frac{b(a)}{a}}-\frac{M_{\rm
shell}}{2a}\right)-\frac{\Lambda}{6}a^3   \,.
\end{equation}

\subsection{Energy conditions at the junction surface}

The junction surface may serve to confine the interior wormhole
exotic matter to a finite region, which in principle may be made
arbitrarily small, and one may impose that the surface
stress-energy tensor obeys the energy conditions at the junction,
$\Sigma$ \cite{VisserBOOK,hawkingellis}.

We shall only consider the weak energy condition (WEC) and the
null energy condition (NEC). The WEC implies $\sigma \geq 0$ and
$\sigma + p \geq 0$, and by continuity implies the null energy
condition (NEC), $\sigma + p\geq 0$.

From eqs. (\ref{surfenergy2})-(\ref{surfpressure2}), we deduce
\begin{equation} \label{sigma+P}
\sigma +p=\frac{1}{8\pi a}
\left[(2-\zeta)\,\sqrt{1-\frac{b(a)}{a}} -
\frac{1-\frac{3M}{a}}{\sqrt{1-\frac{2M}{a}-\frac{\Lambda}{3}a^2}}
\right]  .
\end{equation}
We shall next find domains in which the NEC is satisfied, by
imposing that the surface energy density is non-negative, $\sigma
\geq 0$, i.e., $\sqrt{1-b(a)/a} \geq \sqrt{1-2M/a-\Lambda a^2/3}$.

\subsubsection{1. Schwarzschild solution}

Consider the Schwarzschild solution, $\Lambda=0$ and we impose
that the surface energy density is non-negative, $\sigma \geq 0$.
For the particular case of $\zeta \leq 1$, from equation
(\ref{sigma+P}) we verify that $\sigma+p\geq 0$ is readily
satisfied for $\forall \,a$.

For $1<\zeta <2$, the NEC is verified in the following region
\begin{equation}\label{Schwregion}
2M<a \leq 2M\,\left(\frac{\zeta-\frac{1}{2}}{\zeta-1}\right).
\end{equation}
For convenience, by defining a new parameter $\xi=2M/a$, equation
(\ref{Schwregion}) takes the form
\begin{equation}\label{Schwregion2}
\frac{\zeta-1}{\zeta-\frac{1}{2}} \leq \xi <1 \,.
\end{equation}

For $\zeta=2$, the NEC is satisfied for $\xi \geq 2/3$, i.e., $a
\leq 3M$. For $\zeta > 2$, we need to impose the NEC in the region
of equation (\ref{Schwregion2}); with $\sigma+p<0$ for
$\xi<(\zeta-1)/(\zeta-1/2)$.

\subsubsection{2. Schwarzschild-de Sitter solution}

For the Schwarzschild-de Sitter spacetime, $\Lambda > 0$, we shall
once again impose a non-negative surface energy density, $\sigma
\geq 0$. Consider the definitions $\beta=9 \Lambda M^2$ and
$\xi=2M/a$.

For $\zeta < 2$ the condition $\sigma+p\geq 0$ is readily met for
$\beta \leq \beta_0$, with $\beta_0$ given by
\begin{equation}\label{SdSbeta0}
\beta_0=\frac{27}{4}\,\frac{\xi^2}{(2-\zeta)}\,\left[(1-\zeta)
+\left(\zeta-\frac{1}{2}\right)\xi \right].
\end{equation}
Choosing a particular example, for instance $\zeta=-0.5$, consider
figure 5. The region of interest is shown below the solid line,
which is given by $\beta_r=27\xi^2(1-\xi)/4$. The case of
$\zeta=-0.5$ is depicted as a dashed curve, and the NEC is obeyed
to the right of the latter.

For $\zeta=2$, then the NEC is verified for $\forall \,\beta$ and
$\xi \geq 2/3$, i.e., $r_b<a \leq 3M$, with $r_b$ given by
equation (\ref{root1}). This analysis is depicted in figure 5, to
the right of the dashed curve, represented by $\xi=2$.

For the case of $\zeta>2$, the condition $\sigma+p\geq 0$ needs to
be imposed in the region $\beta_0 \leq \beta \leq \beta_r$; and
$\sigma+p< 0$ for $\beta < \beta_0$. The specific case of
$\zeta=5$ is depicted as a dashed curve in figure 5. The NEC needs
to be imposed to the right of the respective curve.

\begin{figure}[h]
  \centering
  \includegraphics[width=2.2in]{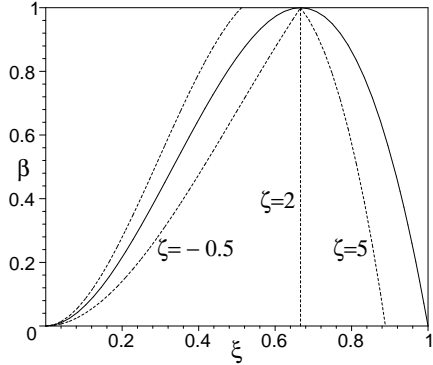}
  \caption{Analysis of the null energy condition for the
  Schwarzschild-de Sitter spacetime. We have considered the
  definitions $\beta=9 \Lambda M^2$ and $\xi=2M/a$.
  Only the region below the solid line is of interest. We have
  considered specific examples, and the NEC is
  obeyed to the right of each respective dashed curves,
  $\zeta=-0.5$,  $\zeta=2$ and $\zeta=5$.
  See text for details.}
\end{figure}

\subsubsection{3. Schwarzschild-anti de Sitter solution}

Considering the Schwarzschild-anti de Sitter spacetime, $\Lambda <
0$, once again a non-negative surface energy density, $\sigma \geq
0$, is imposed. Consider the definitions $\gamma=9 |\Lambda| M^2$
and $\xi=2M/a$.

For $\zeta \leq 1$ the condition $\sigma+p\geq 0$ is readily met
for $\forall \,\gamma$ and $\forall \,\xi$. For $1<\zeta<2$, the
NEC is satisfied in the region $\gamma \geq \gamma_0$, with
$\gamma_0$ given by
\begin{equation}\label{SadSbeta0}
\gamma_0=\frac{27}{4}\,\frac{\xi^2}{(2-\zeta)}\,\left[(1-\zeta)
+\left(\zeta-\frac{1}{2} \right)\xi  \right].
\end{equation}
The particular case of $\zeta=1.8$ is depicted in figure 6. The
region of interest is delimited  by the $\xi$-axis and the area to
the left of the solid curve, which is given by
$\gamma_r=27\xi^2(\xi-1)/4$. Thus, the NEC is obeyed above the
dashed curve represented by the value $\zeta=1.8$.

For $\zeta=2$, then $\sigma+p\geq 0$ is verified for $\forall
\gamma$ and $\xi \geq 3/2$, i.e., $r_b<a \leq 3M$, with $r_b$
given by equation (\ref{adsbhole}). Thus, the NEC is obeyed to the
right of the dashed curve represented by $\zeta=2$, and to the
left of the solid line, $\gamma_r$.

For the case of $\zeta>2$, the condition $\sigma+p\geq 0$ needs to
be imposed in the region $\gamma_r \leq \gamma \leq \gamma_0$. The
specific case of $\zeta=3$ is depicted in figure 6 as a dashed
curve. Thus, the NEC needs to be imposed in the region to the
right of the respective dashed curve and to the left of the solid
line, $\gamma_r$.

\begin{figure}[h]
  \centering
  \includegraphics[width=2.2in]{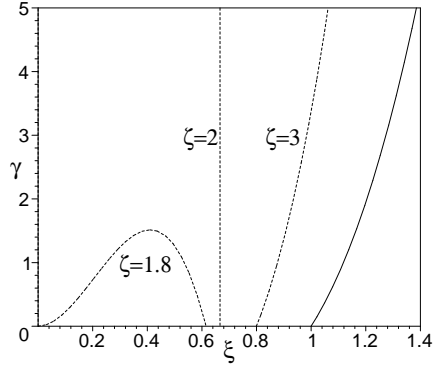}
  \caption{Analysis of the null energy condition for the
  Schwarzschild-anti de Sitter spacetime. We have considered the
  definitions $\gamma=9 |\Lambda| M^2$ and $\xi=2M/a$.
  The only area of interest is depicted to the left of the solid curve,
  given by $\gamma_r=27\xi^2(\xi-1)/4$. For the specific case of
  $\zeta=1.8$, the NEC is obeyed above the respective curve. For
  the cases of $\zeta=2$ and $\zeta=3$, the NEC is verified to the
  right of the respective dashed curves, and to the left of the
  solid line. See text for details.}
\end{figure}

\subsection{Specific cases}

Taking into account equations
(\ref{surfenergy2})-(\ref{surfpressure2}), one may express $p$ as
a function of $\sigma$ by the following relationship
\begin{equation}
p=\frac{1}{8\pi a}
\,\left[\frac{(1-\zeta)+(\zeta-\frac{1}{2})\,\frac{2M}{a}
-(2-\zeta)\frac{\Lambda}{3}a^2}{\sqrt{1-\frac{2M}{a}-
\frac{\Lambda}{3}a^2}} -4\pi a \zeta \sigma \right] .
        \label{Pfunctionsigma}
\end{equation}
We shall analyze equation (\ref{Pfunctionsigma}), namely, find
domains in which $p$ assumes the nature of a tangential surface
pressure, $p>0$, or a tangential surface tension, $p<0$, for the
Schwarzschild case, $\Lambda=0$, the Schwarzschild-de Sitter
spacetime, $\Lambda>0$, and for the Schwarzschild-anti de Sitter
solution, $\Lambda<0$. In the analysis that follows we shall
consider that $M$ is positive, $M>0$.

\subsubsection{1. Schwarzschild spacetime}

For the Schwarzschild spacetime, $\Lambda=0$, equation
(\ref{Pfunctionsigma}) reduces to
\begin{equation}
p=\frac{1}{8\pi a}
\,\left[\frac{(1-\zeta)+(\zeta-\frac{1}{2})\,\frac{2M}{a}}{\sqrt{1-\frac{2M}{a}}}
-4\pi a \zeta \sigma \right] \,.
        \label{SchwarzPfunctionsigma}
\end{equation}
To find domains in which $p$ is a tangential surface pressure,
$p>0$, or a tangential surface tension, $p<0$, it is convenient to
express equation (\ref{SchwarzPfunctionsigma}) in the following
compact form
\begin{equation}
p=\frac{1}{16\pi M} \,
\frac{\Gamma(\xi,\zeta,\mu)}{\sqrt{1-\xi}}\,,
\label{SchwarzcompactP}
\end{equation}
with $\xi=2M/a$ and $\mu=8\pi M\sigma$. $\Gamma(\xi,\zeta,\mu)$ is
defined as
\begin{eqnarray}\label{SchwarzGamma}
\Gamma(\xi,\zeta,\mu)=(1-\zeta)\,\xi+\left(\zeta-\frac{1}{2}\right)\xi^2-\mu
\zeta \sqrt{1-\xi}    \;.
\end{eqnarray}
One may now fix one or several of the parameters and analyze the
sign of $\Gamma(\xi,\zeta,\mu)$, and consequently the sign of $p$.

\bigskip

{\it Fixed $\zeta$, varying $\xi$ and $\mu$.} For instance,
consider a fixed value of $\zeta$, varying the parameters
$(\xi,\mu)$, i.e., consider a fixed value of $a\Phi'(a)$ and vary
the values of the junction radius, $a$, and of the surface energy
density, $\sigma$. It is necessary to separate the cases of
$\zeta=0$, $\zeta>0$ and $\zeta<0$, respectively.

Firstly, for the case of $\zeta=0$, equation (\ref{SchwarzGamma})
reduces to $\Gamma(\xi,\zeta=0,\mu)=\xi-\xi^2/2$, which is always
positive, as $0<\xi<1$, implying a tangential surface pressure,
$p>0$.

Secondly, for $\zeta>0$, the qualitative behavior can be
represented by the specific case of $\zeta=1$, corresponding to a
constant redshift function and depicted in figure 7. For
non-positive values of $\mu$ and $\forall \,\xi$ a surface
pressure, $p>0$, is required to hold the thin shell structure
against collapse. Close to the black hole event horizon, $a
\rightarrow 2M$, i.e. $\xi \rightarrow 1$, a surface pressure is
also needed to hold the structure against collapse. For high
values of $\mu$ and low values of $\xi$, a surface tangential
tension, $p<0$, is needed to hold the structure against expansion.
In particular, for the constant redshift function, $\Phi'(r)=0$,
and a null surface energy density, $\sigma=0$, i.e., $\zeta=1$ and
$\mu=0$, respectively, equation (\ref{SchwarzGamma}) reduces to
$\Gamma(\xi)=\xi^2/2$, from which we readily conclude that $p$ is
non-negative everywhere, tending to zero at infinity, i.e., $\xi
\rightarrow 0$. This is a particular case analyzed in \cite{LLQ}.
Note that a surface boundary, with $p=0$ and $\sigma=0$, is given
by $\xi=(\zeta-1)/(\zeta-1/2)$, for $\zeta>1$.

\begin{figure}[h]
  \centering
  \includegraphics[width=3.0in]{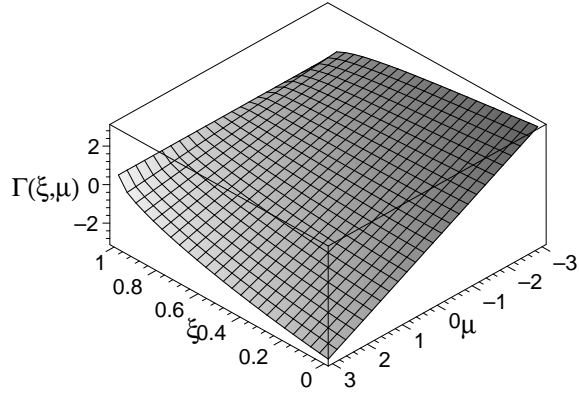}
  \caption{The surface represents the sign of $p$ for the
  Schwarzschild spacetime, $\Lambda=0$, with a
  constant redshift function, $\Phi'(r)=0$, i.e., $\zeta=1$.
  For non-positive values of $\mu$ and $\forall \,\xi$,
  we have a surface tangential
  pressure, $p>0$. For extremely high values of $\xi$
  (close to the black hole event horizon) and $\forall \,\mu$,
  a surface pressure is also required to hold the structure
  against collapse. For high $\mu$ and low $\xi$,
  we have a
  tangential surface tension, $p<0$. See text for details.}
\end{figure}

Finally, for the $\zeta<0$ case, the qualitative behavior can be
represented by the specific case of $\zeta=-1$, depicted in figure
8. For non-negative values of $\mu$ and for $\forall \,\xi$, a
surface pressure, $p>0$, is required. For low negative values of
$\mu$ and for low values of $\xi$, a surface tension is needed,
which is somewhat intuitive as a negative surface energy density
is gravitationally repulsive, requiring a surface tension to hold
the structure against expansion.

\begin{figure}[h]
  \centering
  \includegraphics[width=3.0in]{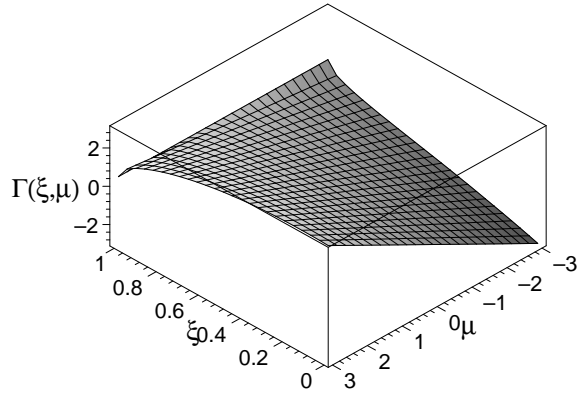}
  \caption{The surface is given by equation (\ref{SchwarzGamma}) for the
  Schwarzschild spacetime, $\Lambda=0$, with $\zeta=-1$.
  For non-negative values of $\mu$ and $\forall \,\xi$,
  we have a surface tangential pressure, $p>0$.
  For negative values of $\mu$ and low $\xi$, a
  tangential surface tension, $p<0$, is required to hold
  the structure against expansion. See text for details.}
\end{figure}

\subsubsection{2. Schwarzschild-de Sitter spacetime}

For the Schwarzschild-de Sitter spacetime with $\Lambda>0$, to
analyze the sign of $p$, it is convenient to express
equation(\ref{Pfunctionsigma}) in the following compact form
\begin{equation}
p=\frac{1}{16\pi M} \,
\frac{\Gamma(\xi,\zeta,\mu,\beta)}{\sqrt{1-\xi-\frac{4\beta}{27\xi^2}}}\,,
\label{compactP}
\end{equation}
with $\xi=2M/a$, $\mu=8\pi M\sigma$ and $\beta=9\Lambda M^2$.
$\Gamma(\xi,\zeta,\mu,\beta)$ is defined as
\begin{eqnarray}\label{GammadS}
\Gamma(\xi,\zeta,\mu,\beta)=(1-\zeta)\,\xi+\left(\zeta-\frac{1}{2}\right)\xi^2
-\frac{4\beta}{27\xi}\,(2-\zeta) -\mu \zeta
\sqrt{1-\xi-\frac{4\beta}{27\xi^2}}    \;.
\end{eqnarray}
One may now fix several of the parameters and analyze the sign of
$\Gamma(\xi,\zeta,\mu,\beta)$, and consequently the sign of $p$.

\bigskip

{\it Null surface energy density.} For instance, consider a null
surface energy density, $\sigma=0$, i.e., $\mu=0$. Thus, equation
(\ref{GammadS}) reduces to
\begin{equation}\label{GammasigmadS}
\Gamma(\xi,\zeta,\beta)=(1-\zeta)\,\xi+\left(\zeta-\frac{1}{2}\right)\,\xi^2-
(2-\zeta)\,\frac{4\beta}{27\xi}\,.
\end{equation}
To analyze the sign of $p$, we shall consider a null tangential
surface pressure, i.e., $\Gamma(\xi,\zeta,\beta)=0$, so that from
equation (\ref{GammasigmadS}) we have the following relationship
\begin{equation}\label{beta0}
\beta_0=\frac{27}{4}\frac{\xi^2}{(2-\zeta)}\left[(1-\zeta)
+\left(\zeta-\frac{1}{2}\right)\,\xi \right]  \,,
\end{equation}
with $\zeta \neq 2$, which is identical to equation
(\ref{SdSbeta0}).

For the particular case of $\zeta=2$, from equation
(\ref{GammasigmadS}), we have
$\Gamma(\xi,\zeta=2,\beta)=\xi(3\xi/2-1)$. A surface boundary,
$\Gamma(\xi,\zeta=2,\beta)=0$, is presented at $\xi=2/3$, i.e.,
$a=3M$. A surface pressure, $\Gamma(\xi,\zeta=2,\beta)>0$, is
given for $\xi>2/3$, i.e., $r_b<a<3M$, and a surface tension,
$\Gamma(\xi,\zeta=2,\beta)<0$, for $\xi<2/3$, i.e., $3M<a<r_c$.

For $\zeta<2$, from equation (\ref{GammasigmadS}), a surface
pressure, $\Gamma(\xi,\zeta,\beta)>0$, is met for $\beta<\beta_0$,
and a surface tension, $\Gamma(\xi,\zeta,\beta)<0$, for
$\beta>\beta_0$. The specific case of a constant redshift
function, i.e., $\zeta=1$, is analyzed in \cite{LLQ} (For this
case, equation (\ref{beta0}) is reduced to $\beta_0=27\xi^3/8$, or
$M=\Lambda a^3/3$). For the qualitative behavior, the reader is
referred to the particular case of $\zeta=-0.5$ provided in Fig.
5. To the right of the curve a surface pressure, $p>0$, is given
and to the left of the respective curve a surface tension, $p<0$.

For $\zeta>2$, from equation (\ref{GammasigmadS}), a surface
pressure, $\Gamma(\xi,\zeta,\beta)>0$, is given for
$\beta_0<\beta<\beta_r$, and a surface tension,
$\Gamma(\xi,\zeta,\beta)<0$, for $\beta<\beta_0$. Once again the
reader is referred to fig. 5 for a qualitative analysis of the
behavior for the particular case of $\zeta=5$. A surface pressure
is given to the right of the curve and a surface tension to the
left.

Note that for the analysis considered in this section, namely, for
a null surface energy density, the WEC, and consequently the NEC,
are satisfied only if $p \geq 0$. The results obtained are
consistent with those of the section regarding the energy
conditions at the junction surface, for the Schwarzschild-de
Sitter spacetime, considered above.

\subsubsection{3. Schwarzschild-anti de Sitter spacetime}

For the Schwarzschild-anti de Sitter spacetime, $\Lambda<0$, to
analyze the sign of $p$, equation (\ref{Pfunctionsigma}) is
expressed as
\begin{equation}
p=\frac{1}{16\pi M} \,
\frac{\Gamma(\xi,\zeta,\mu,\gamma)}{\sqrt{1-\xi+\frac{4\gamma}{27\xi^2}}}\,,
\label{SadScompactP}
\end{equation}
with the parameters given $\xi=2M/a$, $\mu=8\pi M\sigma$ and
$\gamma=9|\Lambda| M^2$, respectively.
$\Gamma(\xi,\zeta,\mu,\gamma)$ is defined as
\begin{eqnarray}\label{GammaSadS}
\Gamma(\xi,\zeta,\mu,\gamma)=(1-\zeta)\,\xi+\left(\zeta-\frac{1}{2}\right)\xi^2
+\frac{4\gamma}{27\xi}\,(2-\zeta) -\mu \zeta
\sqrt{1-\xi+\frac{4\gamma}{27\xi^2}}    \;.
\end{eqnarray}
As in the Schwarzschild-de Sitter solution we shall analyze the
case of a null surface energy density, $\sigma=0$, i.e., $\mu=0$.

\bigskip

{\it Null surface energy density.} For a null surface energy
density, $\sigma=0$, i.e., $\mu=0$, equation (\ref{GammaSadS})
reduces to
\begin{equation}\label{GammasigmaSadS}
\Gamma(\xi,\zeta,\gamma)=(1-\zeta)\,\xi+\left(\zeta-\frac{1}{2}\right)\,\xi^2+
(2-\zeta)\,\frac{4\gamma}{27\xi}\,,
\end{equation}
To analyze the sign of $p$, once again we shall consider a null
tangential surface pressure, i.e., $\Gamma(\xi,\zeta,\beta)=0$, so
that from equation (\ref{GammasigmaSadS}) we have
\begin{equation}\label{SadSbeta}
\gamma_0=\frac{27}{4}\frac{\xi^2}{(2-\zeta)}\left[(\zeta-1)
-\left(\zeta-\frac{1}{2}\right)\,\xi \right]  \,,
\end{equation}
with $\zeta \neq 2$, which is identical to equation
(\ref{SadSbeta0}).

For the particular case of $\zeta=2$, from equation
(\ref{GammasigmaSadS}), we have
$\Gamma(\xi,\zeta=2,\beta)=\xi(3\xi/2-1)$, which is null at
$\xi=2/3$, i.e., $a=3M$. A surface pressure,
$\Gamma(\xi,\zeta=2,\gamma)>0$, is given for $\xi>2/3$, i.e.,
$r_b<a<3M$, and a surface tension, $\Gamma(\xi,\zeta=2,\beta)<0$,
for $\xi<2/3$, i.e., $a>3M$. The reader is referred to the
particular case of $\zeta=2$, depicted in figure 6. A surface
pressure is given to the right of the respective dashed curve, and
a surface tension to the left.

For $\zeta \leq 1$, a surface pressure,
$\Gamma(\xi,\zeta,\gamma)>0$, is given for $\forall \; \gamma$ and
$\forall \;\xi$. For $1<\zeta<2$, a surface pressure,
$\Gamma(\xi,\zeta,\gamma)>0$, is given for $\gamma>\gamma_0>0$;
and a surface tension, $\Gamma(\xi,\zeta,\gamma)<0$, is provided
for $0<\gamma<\gamma_0$. The particular case of $\zeta=1.8$ is
depicted in figure 6, in which a surface pressure is presented
above the respective dashed curve, and a surface tension is
presented in the region delimited by the curve and the $\xi$-axis.

For $\zeta>2$, a surface pressure, $\Gamma(\xi,\zeta,\gamma)>0$,
is met for $\gamma_r<\gamma<\gamma_0$, and a surface tension,
$\Gamma(\xi,\zeta,\gamma)<0$, for $\gamma>\gamma_0$. The specific
case for $\zeta=3$ is depicted in figure 6. A surface pressure is
presented to the right of the respective curve, and a surface
tension to the left.

Once again, the analysis considered in this section is consistent
with the results obtained in the section regarding the energy
conditions at the junction surface, for the Schwarzschild-anti de
Sitter spacetime, considered above. This is due to the fact that
for the specific case of a null surface energy density, the
regions in which the WEC and NEC are satisfied coincide with the
range of $p \geq 0$.

\subsection{Pressure balance equation}

One may obtain an equation governing the behavior of the radial
pressure in terms of the surface stresses at the junction boundary
from the following identity \cite{VisserBOOK,Musgrave}:
$\left[\,T^{\rm
total}_{\hat{\mu}\hat{\nu}}\,n^{\hat{\mu}}n^{\hat{\nu}}
\right]=\frac{1}{2}(K^{i\;+}_{\;\,j} +
K^{i\;-}_{\;\,j})\,S^{j}_{\;\,i}$, where $T^{\rm
total}_{\hat{\mu}\hat{\nu}}=T_{\hat{\mu}\hat{\nu}}-g_{\hat{\mu}\hat{\nu}}\,\Lambda/8\pi$
is the total stress-energy tensor, and the square brackets denotes
the discontinuity across the thin shell, i.e.,
$[X]=X^{+}|_{\Sigma}-X^{-}|_{\Sigma}$. Taking into account the
values of the extrinsic curvatures, eqs.
(\ref{Kplustautau2})-(\ref{Kminustheta2}), and noting that the
tension acting on the shell is by definition the normal component
of the stress-energy tensor,
$-\tau=T_{\hat{\mu}\hat{\nu}}\,n^{\hat{\mu}}n^{\hat{\nu}}$, we
finally have the following pressure balance equation
\begin{eqnarray}\label{pressurebalance}
\left(-\tau^+(a)-\frac{\Lambda ^+}{8\pi} \right) - \left(-\tau
^-(a)-\frac{\Lambda ^-}{8\pi} \right)&=&
\frac{1}{a}\,\left(\sqrt{1-\frac{2M}{a}-\frac{\Lambda}{3}a^2}
+\sqrt{1-\frac{b(a)}{a}}\;\right)\,p
      \nonumber       \\
&&-\left(\frac{\frac{M}{a^2}- \frac{\Lambda}{3}a}
{\sqrt{1-\frac{2M}{a}-\frac{\Lambda}{3}a^2}}+\Phi'(a)\,\sqrt{1-\frac{b(a)}{a}}
\right) \frac{\sigma}{2} ,
\end{eqnarray}
where the $\pm$ superscripts correspond to the exterior and
interior spacetimes, respectively. Equation
(\ref{pressurebalance}) relates the difference of the radial
tension across the shell in terms of a combination of the surface
stresses, $\sigma$ and $p$, given by eqs.
(\ref{surfenergy2})-(\ref{surfpressure2}), respectively, and the
geometrical quantities.

Note that for the exterior vacuum solution we have $\tau^+=0$. For
the particular case of a null surface energy density, $\sigma=0$,
and considering that the interior and exterior cosmological
constants are equal, $\Lambda^-=\Lambda^+$, equation
(\ref{pressurebalance}) reduces to
\begin{equation}
\tau
^-(a)=\frac{2}{a}\,\sqrt{1-\frac{2M}{a}-\frac{\Lambda}{3}a^2}\;\,p
\,.
\end{equation}
For a radial tension, $\tau^-(a)>0$, acting on the shell from the
interior, a tangential surface pressure, $p>0$, is needed to hold
the thin shell form collapsing. For a radial interior pressure,
$\tau^-(a)<0$, then a tangential surface tension, $p<0$, is needed
to hold the structure form expansion.

\subsection{Traversability conditions}

In this section we shall consider the traversability conditions
required for the traversal of a human being through the wormhole,
and consequently determine specific dimensions for the wormhole.
Specific cases for the traversal time and velocity will also be
estimated.

Consider the redshift function given by $\Phi(r)=kr^{\alpha}$,
with $\alpha, k\in \mathbb{R}$. Thus, from the definition of
$\zeta=1+a\Phi'(a)$, the redshift function, in terms of $\zeta$,
takes the following form
\begin{equation}\label{redshift}
\Phi(r)=\frac{\zeta-1}{\alpha}\,\left(\frac{r}{a}\right)^{\alpha}
,
\end{equation}
with $\alpha \neq 0$. With this choice of $\Phi(r)$, $\zeta$ may
also be defined as $\zeta=1+\alpha \Phi(a)$. The case of
$\alpha=0$ corresponds to the constant redshift function, so that
$\zeta=1$. If $\alpha<0$, then $\Phi(r)$ is finite throughout
spacetime and in the absence of an exterior solution we have
$\lim_{r\rightarrow \infty} \Phi(r)\rightarrow 0$. As we are
considering a matching of an interior solution with an exterior
solution at $a$, then it is also possible to consider the
$\alpha>0$ case, imposing that $\Phi(r)$ is finite in the interval
$r_0 \leq r \leq a$.

One of the traversability conditions was that the acceleration
felt by the traveller should not exceed Earth's gravity
\cite{Morris}. Consider an orthonormal basis of the traveller's
proper reference frame, $({\bf e}_{\hat{0}'},{\bf
e}_{\hat{1}'},{\bf e}_{\hat{2}'},{\bf e}_{\hat{3}'})$, given in
terms of the orthonormal basis of the static observers, by a
Lorentz transformation. The traveller's four-acceleration
expressed in his proper reference frame,
$a^{\hat{\mu}'}=U^{\hat{\nu}'} U^{\hat{\mu}'}_{\;\;\;;
\hat{\nu}'}$, yields the following restriction
\begin{equation}\label{travellergravity}
\Bigg| \left(1-\frac{b}{r} \right)^{1/2} \;e^{-\Phi}\,(\gamma
e^{\Phi})'\,c^2 \Bigg| \leq g_{\oplus}  \,,
\end{equation}
where $\gamma=(1-v^2)^{1/2}$, and $v$ being the velocity of the
traveller \cite{Morris}. The condition is immediately satisfied at
the throat, $r_0$. From equation (\ref{travellergravity}), one may
also find an estimate for the junction surface, $a$. Considering
that $(1-b(a)/a)^{1/2}\approx 1$ and $\gamma \approx 1$, i.e., for
low traversal velocities, and taking into account equation
(\ref{redshift}), from equation (\ref{travellergravity}) one
deduces $a \geq |\zeta-1|c^2/g_{\oplus}$. Considering the equality
case, one has
\begin{equation}\label{equalitycase2}
a = \frac{|\zeta-1|c^2}{g_{\oplus}}  \,.
\end{equation}
Providing a value for $|\zeta-1|$, one may find an estimate for
$a$. For instance, considering that $|\zeta-1|\simeq 10^{-10}$,
one finds that $a \approx 10^6 \,{\rm m}$.

Another of the traversability conditions that was required, was
that the tidal accelerations felt by the traveller should not
exceed the Earth's gravitational acceleration. The tidal
acceleration felt by the traveller is given by $\Delta
a^{\hat{\mu}'}=-R^{\hat{\mu}'}_{\;\;\hat{\nu}'\hat{\alpha}'\hat{\beta}'}
\,U^{\hat{\nu}'}\eta^{\hat{\alpha}'}U^{\hat{\beta}'}$, where
$U^{\hat{\mu}'}$ is the traveller's four velocity and
$\eta^{\hat{\alpha}'}$ is the separation between two arbitrary
parts of his body. For simplicity, we shall assume that
$|\eta^{\hat{\alpha}'}|\approx 2\,{\rm m}$ along any spatial
direction in the traveller's reference frame, and that $|\Delta
a^{\hat{\mu}'}|\leq g_{\oplus}$. Consider the radial tidal
constraint,
$|R_{\hat{1}'\hat{0}'\hat{1}'\hat{0}'}|=|R_{\hat{r}\hat{t}\hat{r}\hat{t}}|
\leq g_{\oplus}/2c^2$, which can be regarded as constraining the
metric field $\Phi(r)$, i.e.,
\begin{equation}\label{radialtidalconstraint}
\Bigg|\left(1-\frac{b}{r}\right)
\left(-\Phi''+\frac{b'r-b}{2r(r-b)}\,\Phi'-(\Phi')^2 \right)\Bigg|
\leq \frac{g_{\oplus}}{2c^2}  \,.
\end{equation}
At the throat, $r=r_0$, and taking into account equation
(\ref{redshift}), then equation (\ref{radialtidalconstraint})
reduces to $|(b'-1)\Phi'(r_0)/2r_0| \leq g_{\oplus}/2c^2$ or
\begin{equation}\label{equalitycase}
a =
\left(\frac{|b'-1|\,|\zeta-1|c^2}{g_{\oplus}r_0^2}\right)^{1/\alpha}\;r_0
\,.
\end{equation}
considering the equality case.

Using eqs. (\ref{equalitycase2}) and (\ref{equalitycase}), one may
find an estimate for the throat radius, by providing a specific
value for $\alpha$. Considering $\alpha=-1$ and equating eqs.
(\ref{equalitycase2}) and (\ref{equalitycase}), one finds
\begin{equation}\label{r0}
r_0=\left(\frac{|b'-1|\,|\zeta-1|^2\,c^4}{g_{\oplus}^2}\right)^{1/3}
\,.
\end{equation}
Kuhfittig \cite{Kuhfittig,Kuhfittig2,Kuhfittig3} proposed models
restricting the exotic matter to an arbitrarily thin region under
the condition that $b'(r)$ be close to unity near the throat. If
$b'(r_0)\approx 1$, then the embedding diagram will flare out very
slowly, so that from equation (\ref{r0}), $r_0$ may be made
arbitrarily small. Nevertheless, using the form functions
specified in \cite{Morris} we will consider that $|b'-1|\approx
1$. Using the above value of $|\zeta-1|=10^{-10}$, then from
equation (\ref{r0}) we find $r_0 \simeq 10^4\,{\rm m}$.

To determine the traversal time of the total trip, suppose that
the traveller accelerates at $g_{\oplus}$ halfway to the throat,
then decelerates at the same rate coming to rest at the throat
\cite{Morris,Kuhfittig2}. For simplicity, we shall consider $b'
\approx 1$ near the throat, so the the wormhole will flare out
very slowly. Therefore, from Eq. (\ref{r0}) the throat will be
relatively small, so that we may assume that $r_0 \approx 0$, as
in \cite{Kuhfittig2}. Consider that a space station is situated
just outside the junction surface at $a$, from which the intrepid
traveller will start his/her journey. Thus, the total time of the
trip will be approximately $t_{\rm total} \simeq 4
\sqrt{2a/g_{\oplus}} \approx 1800 \,{\rm s} = 30\,{\rm min}$. The
maximum velocity attained by the traveller will be approximately
$v_{{\rm max}} \simeq 4.5\,{\rm km/s}$.

\section{Conclusion}

By using the cut-and-paste technique we have constructed
thin-shell wormholes in the presence of a generic cosmological
constant. Applying a linearized stability analysis we have found
that for large positive values of $\Lambda$, i.e., the
Schwarzschild-de Sitter solution, the regions of stability
significantly increase relatively to the Schwarzschild case,
analyzed by Poisson and Visser. For negative values of $\Lambda$,
i.e., the Schwarzschild-anti de Sitter, the regions of stability
decrease.

We have also constructed solutions by matching an interior
wormhole spacetime to a vacuum exterior solution at a junction
surface situated at $a$. We have been interested in analyzing the
domains in which the weak and null energy conditions are satisfied
at the junction surface, in the spirit of minimizing the usage of
exotic matter. The characteristics and several physical properties
of the surface stresses were explored, namely, regions where the
sign of tangential surface pressure is positive and negative
(surface tension) were specified. An equation governing the
behavior of the radial pressure across the junction surface was
deduced. Specific dimensions of the wormhole, namely, the throat
radius and the junction interface radius, were found taking into
account the traversability conditions, and estimates for the
traversal time and velocity were also studied.

\begin{acknowledgments}
The author thanks financial support from POSTECH, South Korea, and
CAAUL, Portugal, and thanks Fernanda Fino and the Victor de
Freitas family  for making the trip to South Korea possible.
\end{acknowledgments}

\end{document}